\documentclass[conference]{IEEEtran}
\IEEEoverridecommandlockouts
\usepackage{cite}
\usepackage{amsmath,amssymb,amsfonts}
\usepackage{algorithmic}
\usepackage{graphicx}
\usepackage{textcomp}
\usepackage{booktabs}
\usepackage{xcolor}
\usepackage{url}
\def\BibTeX{{\rm B\kern-.05em{\sc i\kern-.025em b}\kern-.08em
    T\kern-.1667em\lower.7ex\hbox{E}\kern-.125emX}}
\begin{document}

\title{Ensembles at Any Cost? Accuracy-Energy Trade-offs in Recommender Systems}

\author{\IEEEauthorblockN{Jannik Nitschke}
\IEEEauthorblockA{\textit{University of Siegen} \\
Siegen, Germany \\
Jannik.Nitschke@student.uni-siegen.de}
\and
\IEEEauthorblockN{Lukas Wegmeth}
\IEEEauthorblockA{\textit{University of Siegen} \\
Siegen, Germany \\
Lukas.Wegmeth@uni-siegen.de}
\and
\IEEEauthorblockN{Joeran Beel}
\IEEEauthorblockA{\textit{University of Siegen} \\
Siegen, Germany \\
Joeran.Beel@uni-siegen.de}
}

\maketitle

\begin{abstract}
Ensemble methods are frequently used in recommender systems to improve accuracy by combining multiple models. Recent work reports sizable performance gains, but most studies still optimize primarily for accuracy and robustness rather than for energy efficiency. This paper measures accuracy--energy trade-offs of ensemble techniques relative to strong single models.
We run 93 controlled experiments in two pipelines: 1. explicit rating prediction with Surprise (RMSE) and 2. implicit-feedback ranking with LensKit (NDCG@10). We evaluate four datasets ranging from 100{,}000 to 7.8 million interactions (MovieLens-100K, MovieLens-1M, ModCloth, Anime). We compare four ensemble strategies (Average, Weighted, Stacking/Rank Fusion, Top Performers) against baselines and optimized single models. Whole-system energy is measured with EMERS using a smart plug and converted to CO2 equivalents.
Across settings, ensembles improve accuracy by 0.3\% to 5.7\% while increasing energy by 19\% to 2{,}549\%. On MovieLens-1M, a Top-Performers ensemble improves RMSE by 0.96\% at an 18.8\% energy overhead over SVD++. On MovieLens-100K, an averaging ensemble improves NDCG@10 by 5.7\% with 103\% additional energy. On Anime, a Surprise Top-Performers ensemble improves RMSE by 1.2\% but consumes 2{,}005\% more energy (0.21 vs.\ 0.01 Wh), increasing emissions from 2.6 to 53.8 mg CO2 equivalents, and LensKit ensembles fail due to memory limits. Overall, selective ensembles are more energy-efficient than exhaustive averaging, but tuned single models often provide superior accuracy per watt.
\end{abstract}


\section{Introduction}
Recommender systems are deployed at large scale in e-commerce, streaming, and social platforms, where even small increases in compute translate to substantial operational energy use. In parallel, research progress is often reported primarily as accuracy gains, while energy and carbon are rarely disclosed. The Green AI perspective argues that model evaluation should include efficiency metrics alongside predictive performance \cite{schwartz2020greenai,henderson2020systematic}. Recent work has transferred this argument explicitly to recommender systems under the label \emph{Green Recommender Systems} or \emph{Green RecSys} \cite{beel2024greenrecsyscall,wegmeth2025greenrecsys}.

Ensemble techniques combine multiple recommenders to improve prediction quality. In recommender systems research, hybrid and post-hoc ensembles, graph-based combinations, and greedy selection strategies have been reported to yield meaningful improvements in accuracy and robustness \cite{bar2013ensemblecf,boim2011boosting,forouzandeh2021graph,vente2024greedytopn}. At the same time, adjacent AutoML research shows that evaluating and optimizing ensemble construction itself can be computationally expensive enough to motivate dedicated benchmarks and alternative optimization strategies \cite{purucker2022assembled,purucker2023cmaes,purucker2023qdoes}. Yet these benefits come at a cost: ensemble construction and inference multiply training, scoring, and aggregation overhead, potentially increasing the environmental burden in settings where systems are trained repeatedly and served continuously.

This paper quantifies the accuracy--energy trade-offs of common ensemble strategies in recommender systems. We ask: \emph{When do ensembles provide accuracy improvements that justify their energy overhead compared to strong single models?} We make three contributions:
\begin{enumerate}
  \item We provide whole-system energy measurements for ensemble recommender experiments across two pipelines, four datasets, and 93 runs.
  \item We quantify dataset- and strategy-dependent accuracy improvements and energy overheads, showing strongly non-linear trade-offs.
  \item We identify that selective ensembles (Top Performers) are consistently more efficient than exhaustive averaging, and that scalability limits can make ensembles infeasible at larger scale.
\end{enumerate}

\section{Related Work}
\subsection{Ensembles in Recommender Systems}
Ensemble methods in recommender systems include hybrid aggregation, averaging, weighting, stacking, boosting, and rank fusion. Earlier work showed that collaborative-filtering models can be improved through ensemble methods such as boosting and classifier combination \cite{bar2013ensemblecf,boim2011boosting}. More recent recommender-specific work combines graph embeddings with collaborative recommendation and reports substantial gains on MovieLens-style benchmarks \cite{forouzandeh2021graph}. In the most directly related recent work, greedy ensemble selection was adapted to top-$n$ recommendation and consistently outperformed the single best model across five datasets, improving NDCG@10 by 8.6\% on average \cite{vente2024greedytopn}. What this literature largely leaves open is whether these gains remain attractive once the additional energy cost of training and querying multiple recommenders is measured explicitly.

\subsection{Environmental Impact of Recommender Systems}
The environmental impact of recommender systems has only recently become a dedicated research topic. Spillo et al. analyzed the trade-off between recommendation quality and carbon footprint for recommender algorithms \cite{spillo2023sustainability}. Follow-up work studied data reduction as a sustainability lever \cite{spillo2024datareduction}, session-based recommendation under a carbon-performance perspective \cite{plaza2025co2}, graph neural recommenders in an eco-aware setting \cite{purificato2025ecoaware}, and carbon-footprint prediction via meta-modeling \cite{spillo2025carbonator}. Vente et al. provided the first broader paper-level carbon accounting of recommender-systems research by reproducing representative RecSys pipelines \cite{vente2024clickstocarbon}, and this line was later extended into a larger journal article with broader guidance on Green RecSys design and reporting \cite{wegmeth2025greenrecsys}. However, these studies focus on algorithm families, paper pipelines, or sustainability-oriented system design rather than the specific energy cost of ensemble techniques relative to optimized single recommenders.

\subsection{Our Previous Work}
Our earlier work has addressed ensembles from AutoML, recommender-systems automation, and Green RecSys perspectives. In ensemble selection, a first position paper explored whether greedy ensemble selection is feasible for automated recommender systems \cite{vente2022feasibility}. To reduce the cost of comparing ensemble methods, Assembled-OpenML introduced reusable meta-datasets with stored base-model predictions, enabling efficient simulation-based ensemble benchmarks \cite{purucker2022assembled}. Subsequent AutoML work compared greedy ensemble selection to covariance-matrix adaptation for post-hoc ensembling \cite{purucker2023cmaes} and to quality-diversity based population methods \cite{purucker2023qdoes}. Most recently, greedy ensemble selection was adapted directly to top-$n$ recommender evaluation, showing that carefully chosen subsets of classical recommenders can outperform the single best model on ranking tasks \cite{vente2024greedytopn}.

A second line of work in our group concerns automation and algorithm selection as alternatives to combining many models. Recent studies showed that AutoML libraries can perform strongly on recommender-system tasks for inexperienced users \cite{vente2025automl}, and that per-user meta-learning with explicit algorithm characteristics can improve algorithm selection over user-only meta-features \cite{decker2025algoselection}. These works matter for the present paper because they reinforce a key practical baseline: before paying the computational cost of an ensemble, one should ask whether better model selection or automation can deliver similar gains with a single model.

A third line established the Green RecSys perspective. \emph{From Clicks to Carbon} quantified the environmental toll of recommender-systems research and its journal extension generalized this perspective into a broader framework for understanding and minimizing recommender-system carbon footprints \cite{vente2024clickstocarbon,wegmeth2025greenrecsys}. The opinion paper \emph{Green Recommender Systems: A Call for Attention} formulated Green RecSys as a design perspective that explicitly balances performance and environmental cost \cite{beel2024greenrecsyscall}. Methodologically, EMERS introduced hardware-based whole-system energy measurement for recommender experiments \cite{wegmeth2024emers}. Further work investigated dataset downsampling as a way to reduce training cost while preserving recommendation quality \cite{arabzadeh2025dataset}, proposed e-fold cross-validation for recommender-system evaluation \cite{baumgart2025efold}, and explored a broader practical implementation of e-fold cross-validation in machine-learning evaluation \cite{mahlich2024efoldpractice}. The present paper extends this line by focusing specifically on ensembles, for which predictive benefits were known but their energy costs had remained largely unquantified.

\subsection{Energy Measurement Methodologies}
Energy can be approximated via software meters or measured via hardware instrumentation. Software meters are convenient, but they can introduce accuracy and calibration issues depending on hardware support and workload characteristics \cite{jay2023powermeters}. More broadly, Green AI and green AutoML research emphasize that reliable reporting requires transparent measurement assumptions, hardware descriptions, and explicit trade-off analysis \cite{schwartz2020greenai,henderson2020systematic,tornede2023greenautoml}. EMERS operationalizes this recommendation for recommender systems by integrating hardware smart plugs into experiment pipelines and supporting reproducible whole-system measurement \cite{wegmeth2024emers}. We use EMERS to measure complete system energy draw during training and evaluation.

\section{Experimental Design}
\subsection{Datasets and Tasks}
We evaluate explicit rating prediction and implicit-feedback ranking. Table~\ref{tab:datasets} summarizes the datasets. MovieLens-100K and MovieLens-1M are established recommendation benchmarks \cite{harper2015movielens}. ModCloth provides sparse e-commerce interactions, and Anime provides large-scale sparse ratings. Following the bachelor thesis on which this paper is based, preprocessing standardizes all datasets to \texttt{user}, \texttt{item}, and \texttt{rating} columns, removes duplicates and missing values, keeps the most recent interaction when timestamps are available, and removes invalid ratings (including Anime entries coded as ``watched but not rated'').

\begin{table*}[t]
\centering
\caption{Datasets used in the experiments (after preprocessing).}
\label{tab:datasets}
\begin{tabular}{lrrrrr}
\toprule
Dataset & Users & Items & Interactions & Sparsity & Rating scale \\
\midrule
MovieLens-100K & 943 & 1{,}682 & 100{,}000 & 93.70\% & 1--5 \\
MovieLens-1M   & 6{,}040 & 3{,}706 & 1{,}000{,}209 & 95.53\% & 1--5 \\
ModCloth       & 47{,}958 & 1{,}378 & 82{,}790 & 99.87\% & 1--5 \\
Anime          & 73{,}515 & 11{,}200 & 7{,}813{,}737 & 99.05\% & 1--10 \\
\bottomrule
\end{tabular}
\end{table*}

\subsection{Pipelines, Splits, and Metrics}
\textbf{Surprise (explicit rating prediction).}
We use Surprise \cite{hug2020surprise} to predict numerical ratings and report RMSE. Data are split with a global 80/20 random split (fixed seed) and a 5-fold cross-validation on the training set for stability checks.

\textbf{LensKit (implicit-feedback ranking).}
We use LensKit for Python \cite{ekstrand2020lenskit} to generate ranked lists and report NDCG@10 \cite{jarvelin2002cg}. Explicit ratings are converted to implicit positive interactions using thresholding: ratings $\geq 4.0$ for 5-star datasets and $\geq 7.0$ for Anime. Data are split per user (80/20) so each user appears in train and test; users with too few interactions are filtered before splitting to keep the ranking task well-defined on sparse data. We run 5-fold user-based cross-validation on the training split.

\subsection{Models and Ensemble Strategies}
We compare (i) simple baselines, (ii) optimized single models, and (iii) ensembles constructed from the optimized models.

\textbf{Surprise single models.} SVD, SVD++, NMF, KNNBaseline, Slope One, and Co-clustering \cite{hug2020surprise}. The strongest single model is typically SVD++ on MovieLens and ModCloth, and SVD on Anime.

\textbf{LensKit single models.} ALS, BPR, Logistic MF, Item-KNN, and User-KNN \cite{ekstrand2020lenskit}. User-KNN is the strongest single model in our ranking setting for MovieLens and Anime; Item-KNN is competitive on ModCloth.

\textbf{Ensemble strategies.}
We implement four strategies in both pipelines:
\begin{itemize}
  \item \emph{Average:} unweighted average of base model predictions or scores.
  \item \emph{Weighted:} linear combination with weights estimated on validation data.
  \item \emph{Stacking / Rank Fusion:} Surprise uses a linear-regression meta-learner; LensKit uses reciprocal-rank fusion for list aggregation.
  \item \emph{Top Performers:} average of a small validation-selected subset of the best-performing base models; in practice this subset contains only a few models and can differ by dataset depending on feasibility and validation ranking.
\end{itemize}
For rating prediction, the main ensemble candidates are SVD, SVD++, NMF, and KNNBaseline when memory permits; for ranking, the ensemble pool is drawn from the optimized classical recommenders above. This setup mirrors common post-hoc ensembling practice: combine already trained strong models rather than redesign the underlying recommendation algorithms.

\subsection{Energy and Carbon Measurement}
We measure whole-system energy with EMERS \cite{wegmeth2024emers}, using a Shelly Plug smart plug that reports cumulative energy. Each experiment records energy at 0.5\,s intervals. Energy is computed via the delta method:
\begin{equation}
E = E_{\text{end}} - E_{\text{start}}.
\end{equation}
Carbon emissions are estimated from energy using a grid emission factor $EF$:
\begin{equation}
C = (E/1000)\cdot EF,
\end{equation}
where $E$ is in Wh and $EF$ in gCO2e/kWh. We use a European average factor (275\,gCO2e/kWh) as configured in EMERS, enabling direct comparison across runs \cite{wegmeth2024emers}. Because hardware-based smart-plug measurement captures the entire machine, it also captures idle draw and incidental system activity; following EMERS recommendations, runs were therefore executed on an otherwise lightly loaded system.

Experiments were executed on a Windows 11 workstation (AMD Ryzen 7 7800X3D, 32\,GB RAM, AMD Radeon RX 7900 XTX). The measured idle baseline power was 71.2\,W (standard deviation approximately 5\,W), and runs were performed with minimal background load.

\section{Results}
We report results as accuracy change relative to the best optimized single model per dataset and pipeline, and energy overhead relative to that same reference. We focus on representative configurations and summarize additional patterns in text.

\subsection{Explicit Rating Prediction (Surprise)}
Table~\ref{tab:surprise} shows key results for Surprise. On MovieLens-100K, SVD++ is the best single model (RMSE 0.9193 at 0.00182\,Wh), while a Top-Performers ensemble reduces RMSE to 0.9154, corresponding to a 0.42\% improvement at a 52.9\% energy overhead. On MovieLens-1M, Top Performers yields a 0.96\% RMSE reduction over SVD++ with 18.8\% additional energy. This is the most favorable Surprise trade-off in our experiments: the single model is already strong, and the ensemble gains are modest but not entirely dominated by additional energy cost.

On ModCloth, averaging ensembles provide less than 1\% RMSE improvement but increase energy by more than an order of magnitude. The increase is driven by executing multiple base models whose individual costs dominate the small-data workload. On Anime, Top Performers improves RMSE by 1.2\% but requires 0.208\,Wh vs.\ 0.0099\,Wh for SVD (+2{,}005\% energy), increasing emissions from 2.6\,mg to 53.8\,mg CO2e. The Anime setting also revealed a concrete feasibility constraint: KNNBaseline could not be trained because the item-similarity matrix exceeded available memory, which in turn made some ensemble variants impossible.

\begin{table*}[t]
\centering
\caption{Surprise results: best optimized single model vs.\ selected ensembles (energy includes training and evaluation).}
\label{tab:surprise}
\begin{tabular}{llrrlrr}
\toprule
Dataset & Reference single model & RMSE & Energy (Wh) & Ensemble (best-efficiency / best-accuracy) & RMSE & Energy (Wh) \\
\midrule
ML-100K & SVD++ & 0.9193 & 0.00182 & Top Performers (eff.) & 0.9154 & 0.00279 \\
ML-1M   & SVD++ & 0.8625 & 0.03886 & Top Performers (eff.) & 0.8542 & 0.04617 \\
ModCloth & SVD++ & 0.9440 & 0.000117 & Top Performers (eff.) / Average (acc.) & 0.9384 / 0.9357 & 0.000817 / 0.00427 \\
Anime   & SVD   & 1.1335 & 0.00992 & Top Performers (eff.\ \& acc.) & 1.1197 & 0.20842 \\
\bottomrule
\end{tabular}
\end{table*}

\subsection{Implicit-Feedback Ranking (LensKit)}
Table~\ref{tab:lenskit} summarizes LensKit results. On MovieLens-100K, Average improves NDCG@10 by 5.7\% over User-KNN with a 103\% energy overhead. Weighted performs similarly, while Top Performers is cheaper but also less accurate. On MovieLens-1M, reciprocal-rank fusion attains the strongest NDCG@10 but increases energy by roughly 270\% relative to User-KNN; by contrast, ALS is much less accurate but extremely energy-efficient, illustrating that the most sustainable single model need not be the most accurate one.

On ModCloth, the Weighted ensemble reaches the best NDCG@10, but only by 1.3\% over Item-KNN while consuming substantially more energy. On Anime, all ensemble strategies fail due to memory constraints during score aggregation. This indicates that, beyond energy, some ensemble approaches can be infeasible at larger scale without distributed infrastructure or alternative aggregation designs.

\begin{table*}[t]
\centering
\caption{LensKit results: best optimized single model vs.\ selected ensembles.}
\label{tab:lenskit}
\begin{tabular}{llrrlrr}
\toprule
Dataset & Reference single model & NDCG@10 & Energy (Wh) & Ensemble (best-efficiency / best-accuracy) & NDCG@10 & Energy (Wh) \\
\midrule
ML-100K & User-KNN & 0.2176 & 0.000087 & Top Performers (eff.) / Average (acc.) & 0.2238 / 0.2301 & 0.000148 / 0.000177 \\
ML-1M   & User-KNN & 0.1959 & 0.00238 & Average (eff.) / Rank Fusion (acc.) & 0.2028 / 0.2071 & 0.00448 / 0.00889 \\
ModCloth & Item-KNN & 0.1595 & 0.00025 & Top Performers (eff.) / Weighted (acc.) & 0.1603 / 0.1622 & 0.000384 / 0.000441 \\
Anime   & User-KNN & 0.2368 & 0.1777 & (all ensembles failed due to memory limits) & -- & -- \\
\bottomrule
\end{tabular}
\end{table*}

\subsection{Cross-Setting Patterns}
Across 93 runs, ensemble gains are consistently modest relative to their energy overhead. In our setting:
\begin{itemize}
  \item Accuracy improvements range from 0.3\% to 5.7\%, depending on dataset, task, and strategy.
  \item Energy overhead ranges from 19\% (Top Performers on MovieLens-1M in Surprise) to more than 2{,}500\% in settings where the best single model is already energy-efficient and the ensemble executes multiple expensive base models.
\end{itemize}
Two broader patterns stand out. First, on smaller datasets, many models cluster in a relatively narrow accuracy range while still differing strongly in energy consumption. In such settings, even a modest ensemble overhead can dominate the benefit. Second, larger datasets expose wider Pareto fronts, but they also reveal failure modes such as memory exhaustion and overfitting of expensive models. The results therefore align with a diminishing-returns pattern already suggested by Green RecSys work on algorithm-family trade-offs \cite{spillo2023sustainability,vente2024clickstocarbon}.

\section{Discussion}
\subsection{Why the Trade-off is Non-Linear}
The ensemble overhead is not proportional to the marginal accuracy gain. Ensembles incur (i) repeated training across base models, (ii) repeated scoring at inference time, and (iii) aggregation overhead. These costs accumulate even when the base models deliver highly correlated predictions, in which case the additional accuracy gain is limited. The non-linearity is especially visible when the reference single model is already both accurate and cheap, because then even a small absolute increase in energy translates into a very large relative overhead.

\subsection{Selective Ensembles as a Practical Baseline}
Top-Performers ensembles are consistently the most energy-efficient ensemble strategy in our experiments. This supports a practical default: if ensembles are considered, begin with selective combinations of a few strong models rather than exhaustive inclusion. This is consistent with recent work on greedy post-hoc ensembling for top-$n$ recommendation, which also found that a carefully selected subset can outperform any single model \cite{vente2024greedytopn}. It is also consistent with the broader Green RecSys argument that sustainability-aware design should treat performance gains as only one side of the decision problem \cite{beel2024greenrecsyscall,wegmeth2025greenrecsys}.

\subsection{Scalability and Failure Modes}
Beyond energy, memory can be a limiting factor. In LensKit on Anime, ensemble aggregation fails, while single models still execute. In Surprise on Anime, KNN-based training was infeasible because the item-similarity matrix would exceed the available memory budget. These failure modes matter in practice because they make certain ensembles unusable without architectural changes. They also suggest that energy-efficient alternatives such as smaller datasets or fewer evaluation folds \cite{arabzadeh2025dataset,baumgart2025efold,mahlich2024efoldpractice} may be more immediately actionable than adding more models to an already expensive pipeline.

\subsection{Implications for Reporting and Deployment}
For research, reporting accuracy alone is insufficient for comparing methods when computational costs differ substantially. Energy reporting supports more transparent trade-offs and makes work easier to interpret and reproduce \cite{schwartz2020greenai,henderson2020systematic,wegmeth2024emers}. For deployment, the serving cost can dominate: an ensemble that requires multiple model invocations per request increases latency and energy in proportion to traffic volume. Offline or batch settings may tolerate this overhead more easily than real-time serving. Conversely, recent work on AutoML and algorithm selection suggests that stronger automation and better single-model choice may often be a better first step than deploying a fixed ensemble \cite{vente2025automl,decker2025algoselection}.

\section{Limitations}
This study evaluates four classical benchmarks and traditional algorithms; it does not include modern deep learning recommenders. We do not perform systematic hyperparameter optimization across all models, which may shift which single model is ``best'' on a given dataset. Energy is measured as whole-system consumption via a smart plug; this approach captures end-to-end cost but does not separate training from inference and does not attribute energy to individual components (CPU/GPU) \cite{jay2023powermeters,wegmeth2024emers}. Carbon estimates depend on the assumed grid emission factor and should be interpreted as approximations rather than lifecycle assessments.

A second limitation is scope. Our findings concern explicit rating prediction and classical top-$n$ ranking under offline evaluation. They should therefore be read as evidence about the trade-offs of these settings, not as a universal claim about every possible recommender or every possible ensemble. Nevertheless, the methodology is transferable, and recent Green RecSys work suggests that related choices such as data volume, validation design, and algorithm family can already change the sustainability profile of recommender experiments substantially \cite{arabzadeh2025dataset,baumgart2025efold,wegmeth2025greenrecsys}.

\section{Conclusion}
Ensembles in recommender systems can improve accuracy, but the associated energy overhead is often large relative to the gain. Across 93 experiments on two pipelines and four datasets, ensembles improve accuracy by 0.3\% to 5.7\% while increasing energy by 19\% to 2{,}549\%. Selective ensembles (Top Performers) provide better accuracy per watt than exhaustive averaging, yet strong single models (e.g., SVD for rating prediction, ALS or KNN variants for ranking) frequently provide competitive performance at substantially lower energy cost.

These results support three practical recommendations: report energy alongside accuracy in recommender experiments, treat selective ensembles as the default ensemble baseline, and evaluate whether the marginal accuracy gain justifies the operational energy cost for the target deployment setting. More broadly, the results suggest that recommender-systems research should compare ensembles not only against naive baselines, but also against better single-model selection, automation, and Green RecSys design choices \cite{vente2025automl,decker2025algoselection,beel2024greenrecsyscall}.

\section*{Acknowledgment}
This manuscript is based on the bachelor thesis of the first author \cite{nitschke2025thesis}. We used ChatGPT to assist with rewriting and summarizing the bachelor thesis; the authors take full responsibility for the manuscript's content, interpretations, and any potentially remaining errors.

\bibliographystyle{IEEEtran}
\bibliography{references}

\end{document}